\documentclass[sn-nature]{sn-jnl}% Style for submissions to Nature Portfolio journals
\usepackage[LGRgreek]{mathastext}
\usepackage{graphicx}%
\usepackage{multirow}%
\usepackage{amsmath,amssymb,amsfonts}%
\usepackage{amsthm}%
\usepackage{mathrsfs}%
\usepackage[title]{appendix}%
\usepackage{xcolor}%
\usepackage{textcomp}%
\usepackage{manyfoot}%
\usepackage{booktabs}%
\usepackage{algorithm}%
\usepackage{algorithmicx}%
\usepackage{algpseudocode}%
\usepackage{listings}%
\theoremstyle{thmstyleone}%
\theoremstyle{thmstyletwo}%

\theoremstyle{thmstylethree}%

\begin{document}

\title[Article Title]{Near-Petahertz Fieldoscopy of Liquid}
%Near-Petahertz Femtosecond Fieldoscopy of Liquid Samples

\author[1,2]{\fnm{Anchit} \sur{Srivastava}}%\email{iauthor@gmail.com}

\author[1,2]{\fnm{Andreas} \sur{Herbst}}%\email{iiauthor@gmail.com}
\author[1,2]{\fnm{Mahdi M.} \sur{Bidhendi}}%\email{iiiauthor@gmail.com}
\author[1,2]{\fnm{Max} \sur{Kieker}}%\email{iiiauthor@gmail.com}
\author[1]{\fnm{Francesco} \sur{Tani}}%\email{iiiauthor@gmail.com}
\author*[1,2]{\fnm{Hanieh} \sur{Fattahi}}\email{hanieh.fattahi@mpl.mpg.de}

\affil[1]{\orgname{Max Planck Institute for the Science of Light}, \orgaddress{\street{Staudstrasse 2}, \city{Erlangen}, \postcode{91058}, \country{Germany}}}

\affil[2]{\orgdiv{Department of Physics}, \orgname{Friedrich-Alexander-Universit\"at Erlangen-N\"urnberg}, \orgaddress{\street{Staudstrasse 7}, \city{Erlangen}, \postcode{91058}, \country{Germany}}}


\abstract{Measuring transient optical field is pivotal not only for understanding ultrafast phenomena but also for quantitative detection of various molecular species in a sample. In this work, we demonstrate near-petahertz electric field detection of a few femtosecond pulses with 2oo\,attosecond temporal resolution, 10$^8$ detection dynamic range in electric field and sub-femtojoule detection sensitivity, exceeding those reported by the current methods. By field-resolved detection of the impulsively excited molecules in the liquid phase, termed ``femtosecond fieldoscopy'', we demonstrate temporal isolation of the response of the target molecules from those of the environment and the excitation pulse. In a proof-of-concept analysis of aqueous and liquid samples, we demonstrate field-sensitive detection of combination bands of 4.13\,$\mu$mol ethanol for the first time. This method expands the scope of aqueous sample analysis to higher detection sensitivity and dynamic range, while the simultaneous direct measurements of phase and intensity information pave the path towards high-resolution biological spectro-microscopy.}




\maketitle

\section{Introduction}\label{sec1}

Laser-based, label-free quantitative determination of sample composition has proven to be a potent tool across a wide spectrum from fundamental research to real-life applications \cite{han2022understanding, blanco2002nir,espinoza1999characterization,marx2019s, gowen2015feasibility,fiddler2009laser,martin2002near,li2010applications,leighton2022label,krafft2017label,haas2016advances}. For accurate and delicate spectroscopic measurements, it has been crucial to isolate the sample from environmental interferences. For instance, water comprises approximately 60\,\% of the human body, envelops 70\,\% of the Earth's surface, and permeates our surroundings through the atmosphere. Water has a broad absorption spectrum spanning from the visible to mid-infrared (MIR) and is a persistent component on our detectors. Due to its strong absorption cross-section at MIR, sensitive spectroscopy of samples at their resonance frequencies is challenging, as water dominates other, more subtle, absorbance features arising from other molecules. Moreover, the excessive absorbed energy by water at MIR is left in the sample as thermal energy, limiting non-invasive analysis. On the other hand, NIR spectroscopy provides a fingerprint of sample constituents similar to MIR spectroscopy. It distinguishes itself by offering higher spatial resolution and enhanced penetration depth, afforded by the lower absorption cross-section of water in the NIR region \cite{burns2007handbook,bec2019breakthrough,turker2017review}. This feature makes NIR spectroscopy particularly suitable for the non-invasive and label-free examination of soft matter and large-volume aqueous samples \cite{kuda2017determination,bec2020near}.

Overtone and combination vibrations of molecules, which are primarily detected in NIR spectroscopy extend beyond 0.12 petahertz (PHz). Spectrometers have been used for frequency domain detection in this range. However, their detection sensitivity is constrained by the excitation light, which manifests as a background within the same spectral range \cite{mcclure2003204,pasquini2018near,whetsel1968near}. Since the resonance frequencies of overtone and combination bands surpass the sensitivity of silicon-based detectors, employing a second detector, generally more prone to noise in this range, is necessary to capture the system's entire response. Fourier transform spectroscopy is an alternative method allowing for precise spectroscopic detection. Recent advancements in NIR dual-comb spectroscopy have paved the way for precision Fourier transform spectroscopy, albeit primarily in the gas phase \cite{zolot2013broad,zhu2013real,iwakuni2016ortho,coddington2016dual,barreiro2023experimental}. Nonetheless, the technique remains constrained by detectors' spectral response and the existence of a background signal analogous to frequency domain detection \cite{andrews2014quantification,bec2019breakthrough}. 

In contrast, field-resolved detection allows for the direct measurement of light-matter interactions with attosecond precision in a sub-cycle regime, capturing both amplitude and phase information \cite{sommer2016attosecond,riek2015direct}. For decades attosecond streaking was the sole method to probe the electric field of light with a bandwidth approaching the PHz \cite{itatani2002attosecond,goulielmakis2004direct}. A significant drawback was its confinement to vacuum operations. Over the past decade, various techniques have been developed that enable the near-PHz field-resolved detection of light in ambient air \cite{keiber2016electro,cho2019temporal,park2018direct,saito2018all,alismail2020multi,korobenko2020femtosecond,sederberg2020attosecond,sulzer2020determination,zimin2021petahertz,liu2021all,hui2022attosecond,zimin2022ultra,ridente2022electro,altwaijry2023broadband,luo2023synthesis,kempf2023few,yeung2023lightwave}. Among these techniques electro-optic sampling (EOS) stands out for its unparalleled detection sensitivity \cite{herbst2022recent,mamaikin2022contrast,riek2015direct}. In EOS a short probe pulse is employed to resolve the cycles of the electric field of light by up-converting its spectral bandwidth to higher frequencies, making it possible to apply silicon detectors for broadband NIR detection \cite{barh2019parametric}. Moreover, the combination of bright, ultra-short pulses \cite{okamoto2023operation,goncharov2023few,baltuvska1997optical,kottig2020efficient} and heterodyne detection allows for higher detection signal-to-noise ratio and higher detection sensitivity, leaving the shot noise of the probe pulse the primary source of noise \cite{porer2014shot}. 

In this work, we report on the direct detection of the electric field of light at near-PHz frequencies with unparalleled sensitivity and dynamic range. This has been enabled by developing a unique laser source delivering broadband pulses with carrier-to-envelope phase (CEP) stability, which were intrinsically synchronized to near-single-cycle pulses at megahertz (MHz) repetition rates. The unique frontend enabled direct electric-field detection of CEP-stable, few-cycle pulses with unprecedented detection sensitivity and dynamic range via EOS with attosecond temporal resolution. 

Employing the bright ultrashort pulses, we report on the field-sensitive detection of molecular response at the NIR region for the first time. Few-femtosecond, phase-coherent pulses were utilized for both broadband molecular excitation and the near-PHz electric-field detection of their response. Here, the confinement of the excitation pulses allows for temporal gating of the molecular response, while accessing the electric field enables the precise detection of the response of the target molecules from those of the environment. We evaluated our approach by conducting field-resolved detection of water vibration modes in both gas and liquid phases in the NIR region. Additionally, we detected the subtle combination bands of ethanol in the liquid phase. These results show to the best of our knowledge the first field-detection of the NIR molecular response in ambient air, paving the path for the emergence of innovative, field-sensitive, label-free spectroscopy and microscopy techniques. 
\begin{figure}[t!]
\begin{center}
  \includegraphics[width=1\linewidth]{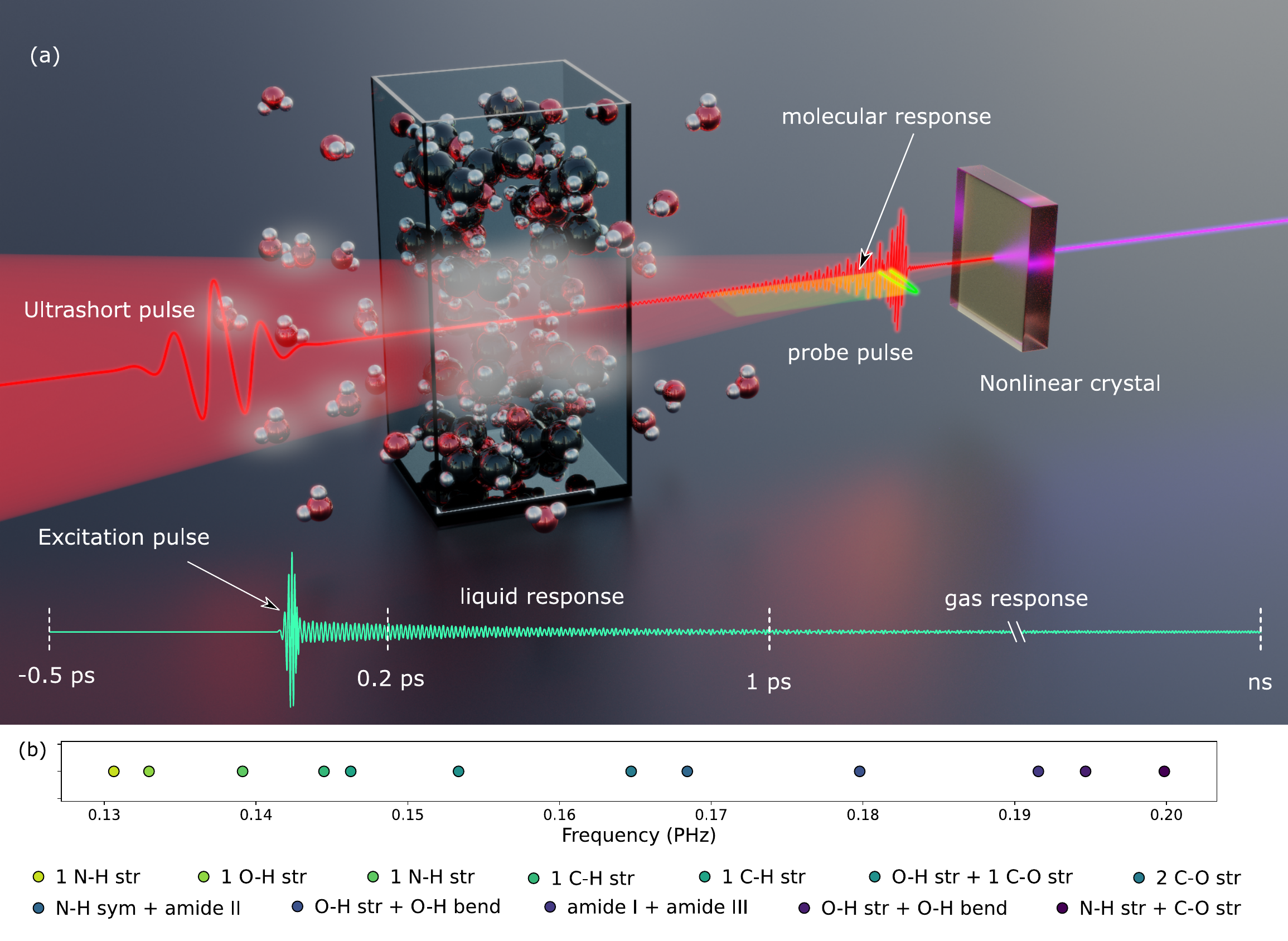}
  \caption{\textbf{Near-Petahertz fieldoscopy.} (a) An ultrashort pulse excites molecules at their NIR resonances. Here, the molecules inside a cuvette represent the sample under scrutiny, while the surrounding molecules represent atmospheric water vapor molecules. The transmitted field contains the global molecular response of both the sample and the environment. A second short pulse at higher frequencies is used for up-conversion and generation of a delay-dependent signal in a nonlinear crystal, where the correlation signal is directly proportional to the electric field of the excitation pulse. The measured electric contains the ultrashort excitation pulse, the delayed response of the liquid spanning over several picoseconds, and a long-lasting response of atmospheric gases lasting for hundreds of nanoseconds. By time filtering and subsequent data analysis, the molecular response can be decomposed to the short-lived liquid and long-lived gas responses. (b) The biological relevant vibrational modes at NIR spectral range. Different compounds like proteins, carbohydrates, lipids, polyphenols, and alcohols are associated with the shown bands. Associated bonds are described in the legend below the plot where ``str'' refers to stretching vibration, whereas ``bend" refers to bending vibration. The numbers one and two indicate the first and second overtones, while the plus sign (+) indicates combination bands. The values were taken from \cite{bec2020near}.}
  \label{fig:fig1}
  \end{center}
\end{figure}

\section{Results}\label{sec2}
Fig.\,\ref{fig:fig1} illustrates the measurement concept. Few-cycle, CEP-stable, NIR pulses, are utilized to extract sensitive spectroscopic information from liquid samples under atmospheric conditions. The electric field of the molecular response, in the wake of the excitation pulse, is resolved via EOS. Due to the high detection sensitivity, we can resolve not only the response of the molecular vibrations of the sample at their overtone and combination resonances but also the response of atmospheric molecules along the beam path. The temporal confinement of the excitation pulses ensures that the molecular responses from both the sample and ambient air are temporally separated from the excitation pulse. Moreover, the response of the liquid sample is temporally distinguished from the ambient air's fingerprint due to the faster dephasing in the liquid phase. Realizing such a concept requires the generation of intrinsically synchronized bright few-cycle pulses with at least one-octave separation and sub-cycle temporal synchronization.\\ 
\begin{figure}[t!]
\begin{center}
  \includegraphics[width=1.05\linewidth]{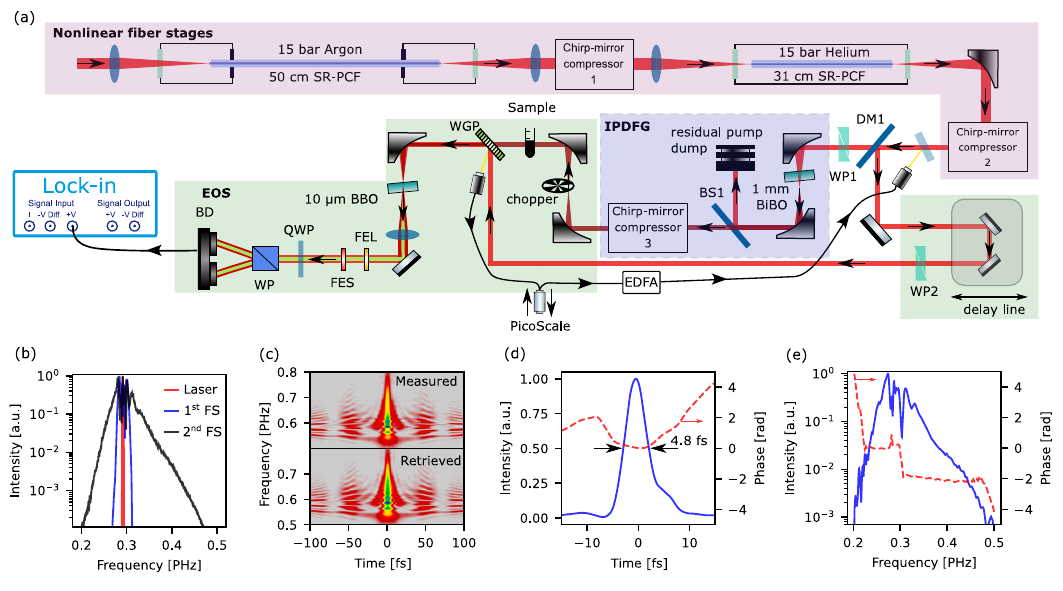}
  \caption{\textbf{Experimental setup.} (a) The shaded regions highlight different parts of the optical setup: nonlinear fiber stages in pink, IPDFG in blue, and EOS in green.
(b) Spectrum of the laser (red), the first fiber stage (blue), and the second fiber stage (black). (c) Measured (top) and retrieved (bottom) spectrograms. The near-single-cycle pulses were measured after the nonlinear compression stages via second-harmonic generation frequency-resolved optical gating (SHG-FROG). (d) The retrieved temporal pulse duration at the output of the second fiber at a 1\,MHz repetition rate. (e) Retrieved spectral intensity and phase. SR-PCF: single-ring hollow-core photonic crystal fiber; DM: dichroic mirror; WP: wedge pair; BS: beam splitter; WGP: wire grid polarizer; FEL: long pass filter; FES: short pass filter; QWP: quarter waveplate; BD: balanced photodiode; BBO: beta barium borate; BiBO: bismuth borate; EDFA: erbium-doped fiber amplifier.}
  \label{fig:fig2}
   \end{center}
\end{figure}
The pink-shaded region in Fig.\,\ref{fig:fig2}(a) shows the optical setup for near single-cycle pulse generation at MHz repetition rates. Single-ring hollow-core photonic crystal fibers (SR-PCF) are used due to their relatively low-loss, broadband guidance, and tunable dispersion \cite{travers2011ultrafast,russell2014hollow}, allowing the generation of ultrashort, bright pulses containing tens of microjoules of energy \cite{kottig2020efficient,schade2021scaling}. In the first stage, 20\,$\mu$J, 1\,MHz laser pulses were compressed from 255\,fs to 25\,fs at full width at half maximum (FWHM), by self-phase-modulation based spectral broadening in Argon-filled SR-PCF (see supplementary information (SI) Fig.\,\ref{fig:FS} and Fig.\,\ref{fig:FD}), followed by group-delay dispersion compensation by a chirped mirror (CM) compressor. Subsequently, in the second stage, the 25\,fs pulses were compressed to a near single-cycle duration via soliton-effect self-compression in a similar fiber (Fig.\,\ref{fig:fig2}(b)). In both stages, the gas species used as a medium were selected to minimize photoionization and subsequent long-lived effects occurring at MHz repetition rates \cite{koehler2021post}. The accumulated dispersion on the near single-cycle pulses due to propagation in media after the fiber was compensated in a CM compressor to 4.8\,fs at FWHM, limited by the bandwidth of the CM compressor (Fig.\,\ref{fig:fig2}(c) and Fig.\,\ref{fig:fig2}(d)). This corresponds to 3.5\,GW of peak power with 69\,\% of the energy in the main pulse. The retrieved spectrum shown in Fig.\,\ref{fig:fig2}(e) spans over 300\,THz bandwidth supporting 3\,fs pulses at FWHM.
\begin{figure}[t!]
  \includegraphics[width=\linewidth]{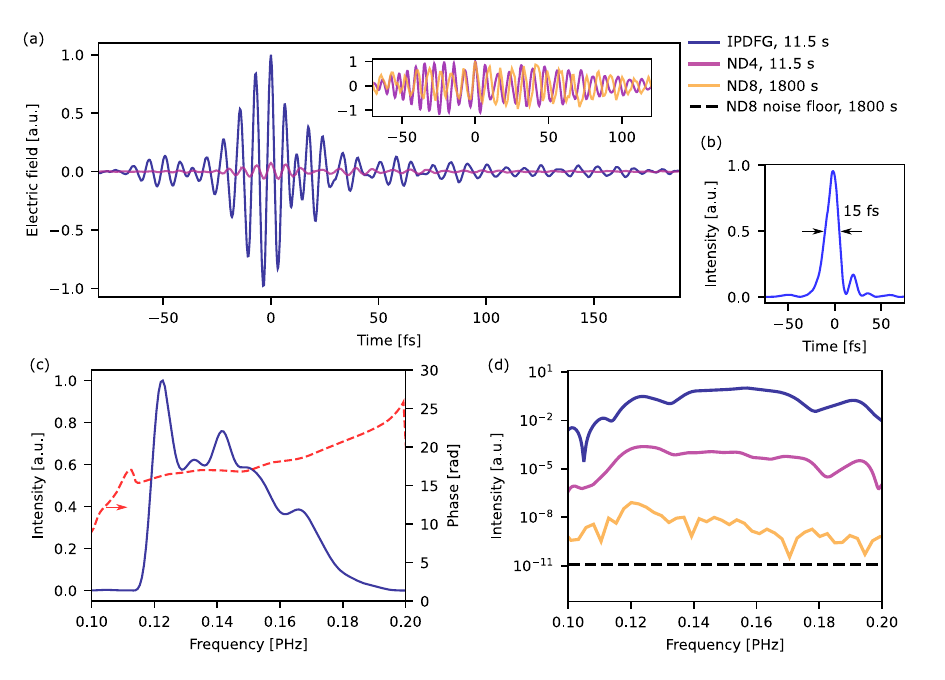}
  \caption{\textbf{NIR electric field sampling.} (a) The measured electric field of CEP-stable pulses via EOS. The inset displays two low-energy fields in the presence of ND4 (magenta) and ND8 filters (yellow). Both fields are longer than the blue curve due to additional material dispersion caused by the ND filters. (b) The temporal profile of the CEP-stable pulses with a pulse duration of 15\,fs FWHM. (c) Retrieved spectrum and phase of the CEP-stable pulses. (d) Retrieved spectrum of the CEP-stable pulses after attenuation with the ND4 and ND8 filters. The spectrum of the unattenuated pulse is shown in blue for comparison. The dashed black line represents the measured noise floor in the absence of the CEP-stable pulses on the balanced detector. The legend displays the acquisition time for each field, along with the ND filter label. This highlights the system's ability to measure fields with a dynamic range greater than 10$^8$ in the electric field. ND: neutral density. The retrieved spectra are not corrected for the spectral response of the filters.}
  \label{fig:fig3}
\end{figure}

Intrapulse difference frequency generation (IPDFG) was employed to generate broadband NIR pulses with a passive CEP-stability (blue region in Fig.\,\ref{fig:fig2}(a)) \cite{fattahi2013efficient}. A 500\,$\mu$m-thick bismuth borate (BiBO) crystal was pumped by 4.8\,fs pulses at 1\,$\mu$m to generate a broadband spectrum spanning from 0.1\,PHz to 0.23\,PHz and 160\,mW of average power. A custom-made dichroic beam splitter with the cut-off at 0.2\,PHz was used to separate the residual pump from the CEP-stable pulses. After dispersion management with a custom-made CM compressor, the NIR pulses along with the fraction of 4.8\,fs pulses were sent to a 20\,$\mu$m beta barium borate (BBO) crystal for electric-field sampling. The green-shaded region in Fig.\,\ref{fig:fig2}(a) highlights the schematic of the NIR EOS. The up-converted signal is spectrally filtered with a pass band filter from 0.425\,PHz to 0.5\,PHz to enhance the detection sensitivity by eliminating the spectral components that do not carry specific field information. The measured electric field of the broadband CEP-stable pulses and its temporal intensity profile with a 15\,fs pulse duration at FWHM are shown in Fig.\,\ref{fig:fig3}(a) and Fig.\,\ref{fig:fig3}(b), respectively. The corresponding spectral intensity and phase obtained through the Fourier transform shown in Fig.\,\ref{fig:fig3}(c) reveals residual higher-order dispersion, which can be compensated by optimizing the design of the CM compressor. While the high-frequency cut-off of the spectrum is limited by the beam splitter roll-off, the crystal absorption constrains the low-frequency cut-off to 0.1\,PHz. To verify the detection sensitivity and dynamic range of the detector, the energy of the IPDFG pulses before EOS crystal was reduced from 80\,nJ to sub-femto joule energies by using a series of neutral density filters. Fig.\,\ref{fig:fig3}(a) inset shows two field-resolved measurements at 8\,pJ (magenta) and 0.7\,fJ (in yellow), respectively. The spectral intensity counterpart at the three different pulse energies is shown in Fig.\,\ref{fig:fig3}(d) corresponding to a 110 dB dynamic range. To establish the ability of the system to detect the response of minute quantities of molecules, we resolved the electric field of the atmospheric water vapor molecules, liquid water, and ethanol in ambient air after excitation by femtosecond NIR pulses. The temporally gated molecular response and their frequency counterparts are shown in Fig.\,\ref{fig:fig4}.

\section{Discussion}\label{sec2}
\begin{figure}[t!]
  \includegraphics[width=\linewidth]{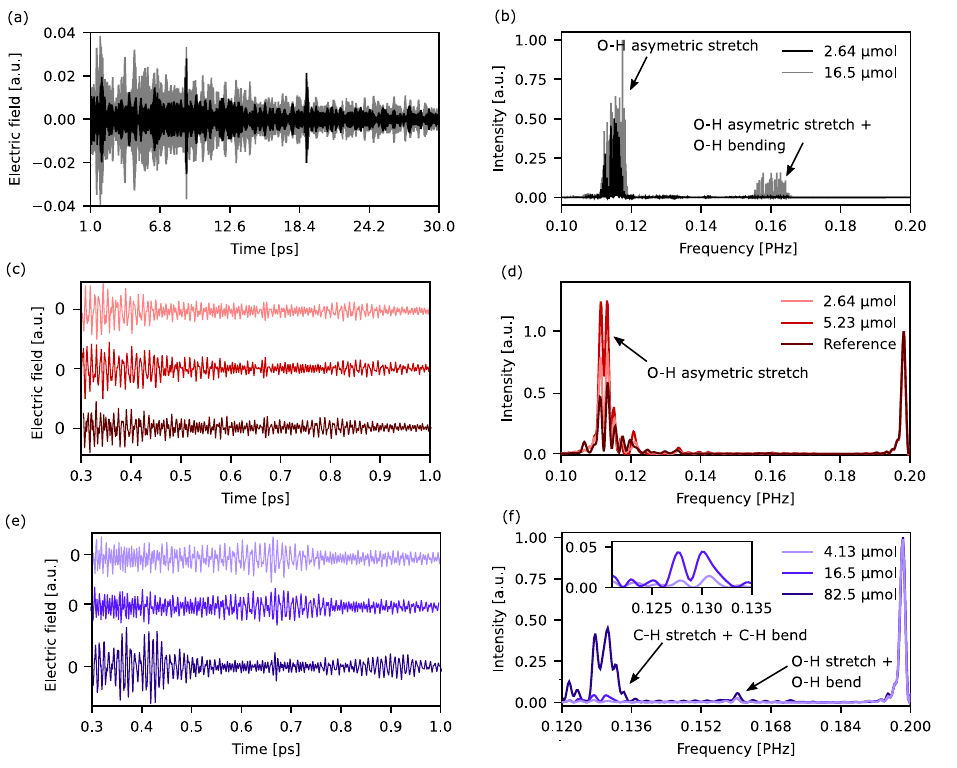}
  \caption{\textbf{Benchmarking measurements.} Each row shows the time-gated electric field on the left and its Fourier-transformed spectrum on the right. (a) The light grey field represents atmospheric water molecules at 50\,\% RH (16.5 $\mu$mol), while the black curve represents 8\,\% RH (2.64 $\mu$mol). (b) Two absorption modes are visible: a fundamental mode at 0.115\,PHz and a combination band at 0.16\,PHz. (c) Liquid water molecules at 5.23\,$\mu$mol (red) and 2.64\,$\mu$mol (pink) along with pure acetic acid (brown) as a reference measurement. (d) The molecular response of pure acetic acid is compared to that of aqueous solutions. (e) Pure liquid ethanol at different volumes of 4.13\, $\mu$mol, 16.5\,$\mu$mol, and 82.5\,$\mu$mol. (f) The weak combination peak centered at 0.130\,PHz is observed at all three volumes.} 
  \label{fig:fig4}
\end{figure}
Water has two prominent absorption peaks within the spectral coverage of the excitation pulses: i) an asymmetric stretch centered at 0.115\,PHz (3836\,cm$^{-1}$) and ii) a combination resonance of bending and asymmetrical stretch centered at 0.16\,PHz (5337\,cm$^{-1}$) \cite{Levine_1975}. Fig.\,\ref{fig:fig4}(a) shows the electric field of atmospheric water vapor molecules at two different laboratory relative humidity (RH). The measurements at 50\,\% RH and 8\,\% RH, correspond to 16.5\,$\mu$mol and 2.64\,$\mu$mol of atmospheric water vapour molecules interacting with the broadband 15\,fs excitation pulses, respectively (see SI). The minimum laboratory achievable RH of 8\,\% was reached by purging the beam path with dry air and nitrogen. Fig.\,\ref{fig:fig4}(b) shows the corresponding absorption frequencies for both concentrations. The spectra are achieved by Fourier-transformation of the temporally gated molecular response at 1\,ps after the excitation pulse for a temporal window of 30\,ps and agree with the HITRAN database  \cite{gordon2022hitran2020} (see SI Fig.\ref{fig:HITRAN}). At the 2.64\,$\mu$mol level, the absorption peak at 0.16\,PHz is barely resolved due to the low absorption cross-section of the water's combination band (65$\times$10$^{-21}$cm$^2$/molecule) compared to its counterpart at the fundamental resonance (600$\times$10$^{-21}$cm$^2$/molecule) \cite{gordon2022hitran2020}.

Fig.\,\ref{fig:fig4}(c) shows the measured molecular response of 5.23\,$\mu$mol and 2.64\,$\mu$mol diluted water in the liquid phase, which were prepared by mixing 20\,$\mu$L and 10\,$\mu$L of deionized water in 1\,mL acetic acid. We examined the molecular response in the liquid phase in the temporal window of 0.3\,ps to 1\,ps, as dephasing occurs faster (Fig.\,\ref{fig:fig4}(d)) \cite{falk1984frequency,bruenig2022time}. The absorption amplitude of both concentrations was normalized to 0.2\,PHz peak, which is present at this time scale due to the cut-off of the dichroic beam splitter used in our setup. Water's asymmetric stretch resonance at 0.115\,PHz (3836 $cm^{-1}$) in Fig.\,\ref{fig:fig4}(d) is distinct for 5.23\,$\mu$mol (red curve) and 2.64\,$\mu$mol (magenta curve). The pure acetic acid response in the presence of water vapor molecules at RH of 8\,\% is shown in Fig.\,\ref{fig:fig4}(d) (black curve). To evaluate the detection sensitivity of our setup in environmental conditions, ethanol was measured due to its distinct resonance frequency in comparison to water and its low absorption cross-section. Fig.\,\ref{fig:fig4}(e) and Fig.\,\ref{fig:fig4}(f) show the resonance of pure liquid ethanol at different concentrations and 8\,\% RH, in time and frequency domain, respectively. The weak absorption at 0.13\,PHz (4336 $cm^{-1}$) is due to the combination band resonance arising from C-H stretch and C-H bend mode, which was resolved very clearly in our measurement at the minimum detectable amount of 4.13\,$\mu$mol \cite{davies1988identification}.

We define a figure of merit (FOM) to allow detection sensitivity comparison of absorption bands between different species and various absorption cross-sections. The FOM is defined as:
\begin{equation}
    FOM = n \times \sigma
\end{equation}
%FOM value for waters fundamental vibration at 2.5um is 1.584x10^-28
Where n is the amount of substance in mol and $\sigma$ is the absorption cross-section of the absorption band in cm$^2$/molecule. Given the absorption cross-section values of water (600$\times$10$^{-21}$\,cm$^2$/molecule) and ethanol (3.2$\times$10$^{-21}$\,cm$^2$/molecule) \cite{gordon2022hitran2020,sharpe2004gas}, the calculated FOM value for water at 2.64\,$\mu$mol is $9\times10^{-28}$\,m$^2$\,mol/molecule, while FOM for ethanol's combination band at 4.13\,$\mu$mol is $1.322\times10^{-30}$\,m$^2$\,mol/molecule with merely 25\,$\mu$m path length. 

In conclusion, we report on field-sensitive, near-PHz detection of molecular fingerprints in the liquid phase for the first time. To this end bright, intrinsically synchronized CEP-stable, 15\,fs pulses at 2\,$\mu$m, and 4.8\,fs pulses at 1\,$\mu$m were generated for impulsive excitation and probing of the molecular response via EOS. Using the MHz ultrashort laser pulses, we demonstrated the ambient air field-resolved detection of femtosecond pulses with sub-femtojoule energy, and $10^{8}$ detection dynamic range in the electric field, enhancing the near-PHz field-detection sensitivity by three orders of magnitude compared to other field-resolved techniques \cite{herbst2022recent}. The source’s MHz repetition rate augments both the signal-to-noise ratio and detection sensitivity, while the signal upconversion in our scheme alleviates the bandwidth constraint inherent in silicon-based detectors. In a proof of concept, we reported on the sensitive detection of the vibration modes of atmospheric and aqueous water molecules at 2.64\,$\mu$mol, and ethanol combination band at 4.13\,$\mu$mol. To the best of our understanding, these measurements mark the first field-resolved detection of both fundamental and combination bands in the liquid phase. 

Femtosecond pump-probe spectroscopy has provided evidence that rapid dynamics occurring within a time scale of fewer than 100\,fs of liquid water have a significant impact on chemical reactions taking place in the aqueous phase \cite{jimenez1994femtosecond}. This underscores the vital importance of ultrafast processes in comprehending aqueous phase chemistry for example on grasping how water molecules dissipate energy \cite{perakis2016vibrational}. The electric field measurements presented in Fig.\,\ref{fig:fig4} thus establish a foundation for studying aqueous solutions with enhanced sensitivity and dynamic range compared to femtosecond intensity pump-probe techniques \cite{maiuri2019ultrafast, crowell1995infrared}, rooted in the higher amplitude of the molecular response relative to the excitation pulses in field-resolved detection (see SI Fig.\,\ref{fig:IF}) \cite{pupeza2020field,kowligy2019infrared}. Furthermore, the high repetition rate of near-single-cycle pulses not only lays the groundwork for single-shot monitoring of chemical reactions in liquids \cite{couture2023single} but also presents intriguing possibilities for exploring nonlinear interactions due to the unique combination of peak and average power in the near-single-cycle domain. 

Stimulating the molecular composition of a sample with phase-coherent femtosecond excitation pulses leads to temporal gating between the molecular response from the excitation pulses. Moreover, accessing the sub-cycle electric field of light allows for decomposing the short-lived liquid molecular response from the long-lived ambient gas responses (see SI Fig. \ref{fig:TG}). Femtosecond fieldoscopy expands the scope of aqueous sample analysis and paves the path toward novel methods for multi-dimensional spectroscopy and high-resolution biological spectro-microscopy.

\section{Materials and Methods}
\subsection{Near single-cycle pulse generation}
A commercially available Yb:KGW amplifier (CARBIDE Light Conversion) delivering 255\,fs pulses at 1030\,nm with 20\,W of average power, and at 1 MHz repetition rate, is used as the source laser. In the first nonlinear fiber stage, we used a 100\,mm focal length (Thorlabs LA1509-B-ML) lens to couple 20\,$\mu$J circularly polarized pulses into a 50 cm long SR-PCF with a core diameter of 55\,$\mu$m (see SI Fig.\,\ref{fig:SEM}) filled with 15\,bar of Argon. The spectrally broadened laser pulses were compressed to 25\,fs (FWHM) using a CM compressor (UltraFast Innovations GmBH PC1611) with 12 bounces (see SI Fig.\,\ref{fig:S1}). The total group delay dispersion compensated by the CM compressor was -1800\,fs$^2$. Subsequently, the 25\,fs compressed pulses were coupled (via Thorlabs LA1509-BML) to a 31\,cm long SR-PCF (parameter same as before), filled with 15\,bar of Helium. After the second stage, a 2-inch off-axis silver parabola (Edmund 36-598) with an effective focal length of 177.8\,mm was used to collimate the beam. The two gas cells were mounted on a 3-axis stage (MDE122) from Elliot Scientific. The two-stage fiber system achieves a total throughput of 85\,\% before the collimating parabola. Afterward, the soliton-compressed pulses were sent to a CM compressor (UFI PC105) consisting of four double-angle CM (total group delay dispersion of -160\,fs$^2$) and a pair of wedges (Altechna M0067705) to compensate for the dispersion caused by the two mm-thick MgF$_2$ output window of the second gas cell and to pre-compensate for the accumulated dispersion on the pulse before reaching the IPDFG crystal. After the compressor, we placed a 1 mm AR-coated window (UFI AR7203) that reflects 5\,\% of the beam, to separate the probe pulses for EOS. We employed a home-built second-harmonic generation frequency-resolved optical gating (SHG-FROG) for temporal characterization. To ensure accurate measurements, the device utilized all-reflective dispersion-free optics in a non-collinear geometry with 10\,$\mu$m-thick BBO crystal (Castech) cut for type I phase matching. 

\subsection{CEP-stable, NIR pulse generation}
For the IPDFG, we focused 12\,$\mu$J of the compressed pulses from the fiber stages to 40\,$\mu$m by using a 1-inch off-axis parabola. A type II BiBO crystal with a phase matching angle of 12°\,(Castech) was placed a few millimeters behind the focus to avoid white light generation in the crystal. A half-wave plate was introduced into the beam path before the second fiber stage to rotate 1\,\% of the input `p' polarisation state to `s' polarisation state. The generated CEP-stable pulse was collimated to a beam size of 3.2\,mm at 1/e$^2$ employing a 4-inch focal length parabola (Thorlabs MPD254508-90-P01). A custom-built (UFI BS2214-RC2) broadband dichroic beam splitter separated the pump and the NIR beam. The measured power after the beam splitter was 160\,mW. A custom-built double-angle CM compressor (UFI IR7202) (see SI Fig.\,\ref{fig:CM}) with four bounces was used for temporal compression to 15\,fs.

\subsection{Field-resolved detection}
In EOS, probe and excitation pulses propagate collinearly in a nonlinear crystal, generating spectral components at sum and difference frequencies. The up-converted field-sensitive signal arises from the interference between partially overlapping spectra of the probe pulse and the sum or difference frequency pulse. Through an ellipsometer, direct access is obtained to the electric field of the sampled pulse \cite{sulzer2020determination}. By utilizing a lock-in amplifier and balanced detection, the technical noise surplus of the gate pulse was mitigated, thereby making the shot noise of the probe pulse the primary limitation on detection sensitivity. In the EOS, the IPDFG pulses were used as an excitation pulse, whereas a 5\,\% reflection from the second fiber stage’s output was used as a probe pulse. A wire grid polarizer combined the probe and the excitation pulses with orthogonal polarization, collinearly in the EOS crystal. An off-axis parabolic mirror of 3-inch focal length was used to focus the beams in a 20-$\mu$m-thick, type II BBO crystal to generate the sum frequency signal. The sum frequency signal interferes with the high-frequency portion of the probe pulse, which acts as a local oscillator for a heterodyne detection. Appropriate filters (Thorlabs FEL 600 and FES 700) were placed after the EOS crystal. The resulting polarization rotation was measured by an ellipsometer, which included a Wollaston prism, a quarter waveplate, and balanced photodiodes. The quarter-wave plate is adjusted to ensure that both photodiodes receive the same intensity in the presence of the probe pulses. A mechanical chopper modulated the excitation pulses at 5.8 kHz to enable heterodyne lock-in detection. The delay line was based on the linear motorized stage (Physik Instrument V-528.1AA) with a scanning range of 20\,mm, corresponding to a scanning delay of 132\,ps. An interferometric delay tracking system \cite{schweinberger2019interferometric} was employed to precisely track the delay line and any timing jitter artifacts (see SI).

In EOS, the measured interference of the sampling field with sum frequency field components is convoluted with the detector response. Consequently, it is subject to a complex response function comprising both amplitude and phase components. The response function can be calculated using the methods described in \cite{keiber2016electro}. Based on the wavevector mismatch calculation (see SI Fig.\,\ref{fig:MM}), it can be seen that the post-processing of the measured field can be neglected for our spectral range as the nonlinear response remains constant throughout the EOS detection range. 

\subsection{Sample preparation} 
A Pike Technologies liquid cell (162-1200) with two 3\,mm-thick barium fluoride windows and a spacer was used to hold the liquid samples. To examine the liquid water, 10\,$\mu$L and 20\,$\mu$L of deionized water were mixed with 1\,mL acetic acid buffer solution (chemlab - CL00.0119). The samples were placed between two windows using a 0.5 mm Teflon O-ring, corresponding to an irradiation volume of 4.81\,$\mu$L. Three cells with different concentrations were mounted side to side on a translation stage to reduce the systematic error. For ethanol measurements, pure ethanol (VWR chemicals - 85033.360) was filled into three liquid cells with Teflon O-ring spacers of different thicknesses (0.5\,mm, 0.1\,mm, and 0.025\,mm). The spacers corresponded to irradiation volumes of 4.81\,$\mu$L, 0.962\,$\mu$L, and 0.241\,$\mu$L, respectively.

\section*{Acknowledgements}
We thank PSJ Russell, MH Frosz, and their team for the production of the fibers used in this experiment. We want to express our deep gratitude to Daniel Schade, Wolfgang Schweinberger, Gunnar Arisholm, Nicholas Karpowicz, and Mallika I Suresh for their invaluable guidance and support.
\section*{Declarations}
\begin{itemize}
    \item This work was supported by research funding from the Max Planck Society.
    \item Conflict of interest/Competing interests: The authors do not declare any competing interests.
    \item Authors' contribution: H.F. envisioned and designed the experiment. A.S., A.H., F.T., M.K. implemented fiber stages for short pulse generation. A.S, A.H. M.B implemented the data acquisition system. A.S. performed the fieldoscopy measurements. A.S. and H.F performed the data analysis and wrote the manuscript. All authors proofread the manuscript.
\end{itemize}

\clearpage

\section*{Supplementary Information}
\subsection*{Data acquisition} \label{acquistion}
The experimental configuration incorporates custom-developed software tailored to comprehensively control various components involved in the measurement process, namely the delay stage, PicoScale, and lock-in amplifier. The delay stage employed is a Physik Instrument V-528.1AA, specifically chosen for its suitability within the experimental framework. The PicoScale, an integral setup component, is a commercially available interferometer manufactured by SmarAct. It operates on the principles of sinusoidal phase modulation for precise interferometric displacement measurements. Notably, the PicoScale measures temporal delays and detects any mechanical jitter introduced within the beam path. Integrating the interferometer into the experimental arrangement involves amplifying the PicoScale laser (1550\,nm, CW) through an erbium-doped fiber amplifier and directing it parallel to the optical setup. This enables accurate tracking of the optical path difference between the pulses with attosecond precision. To facilitate lock-in measurements, an optical chopper modulates the NIR beam at a frequency of 5.882\,kHz. For the presented measurements, the stage velocity is 0.4\,mm/s. 

\subsection*{Concentration Calculations} \label{cal}
The number of molecules is calculated by the ideal gas law:
\begin{equation}\label{gaslaw}
    PV = nRT
\end{equation}
where, P = pressure, V = volume, n = Number of moles, R = ideal gas constant (8.314 J/mol K) and T = temperature in Kelvin. 3.2\,m of beam propagation in the air with a beam diameter of 3.5\,mm corresponds to a volume of 3.08\,$\times$ 10$^{-5}$\,m$^3$ (0.0308\,L). At 22°\,C (295.15\,K), the saturation vapor pressure of water ($P_{saturated}$) is 2.64\,kPa. To calculate the partial pressure of water vapor $P_{water}$, we use:
\begin{equation}
    P_{water} = RH\,\times\,P_{saturated}
\end{equation}
For RHs of 50\,\% and 8\,\%, the corresponding values for $P_{water}$ are 1.32\,kPa and 0.211\,KPa respectively. Using the equation \ref{gaslaw}, we have $n_{RH50\,\%} = 16.5\,\mu$mol and $n_{RH8\%} = 2.64\,\mu$mol. 
 \\
For liquid water samples, we dissolved 10\,$\mu L$ and 20\,$\mu$L of deionized water in 1\,mL of acetic acid buffer, corresponding to a Molarity of 0.555\,M and 1.088\,M, respectively (taking into account the density of water as 1\,g/mL and the molar mass of water as 18.02 g/mol). For a beam diameter of 3.5\,mm and an O-ring spacer of 0.5\,mm, irradiated volume is calculated to be 4.81\,$\mu$L. To determine the number of moles in the irradiation volume, the irradiation volume is multiplied by the molarity of the corresponding solution. For 10\,$\mu L$ and 20\,$\mu L$  dissolved water, we obtain 2.64\,$\mu$mol and 5.23\,$\mu$mol, respectively.

For calculating the number of moles of ethanol at three different volumes of 4.81\,$\mu L$, 0.55\,$\mu L$, and 0.137\,$\mu L$: firstly the corresponding mass was calculated by multiplying the volume and density. Afterward, the obtained values were divided by ethanol's molar mass. Obtaining 0.79\,g/ml as the density of ethanol, and 46.068\,g/mol for its molar mass, the number of moles was calculated to be 82.5\,$\mu$mol, 16.5\,$\mu$mol, and 4.13\,$\mu$mol, respectively. Molar mass and density information for all compounds were retrieved from PubChem.\cite{kim2023pubchem}.

\clearpage

\subsection*{Nonlinear propagation in both fibers}

\begin{figure}[h]
\begin{center}
  \includegraphics[width=1.1\linewidth]{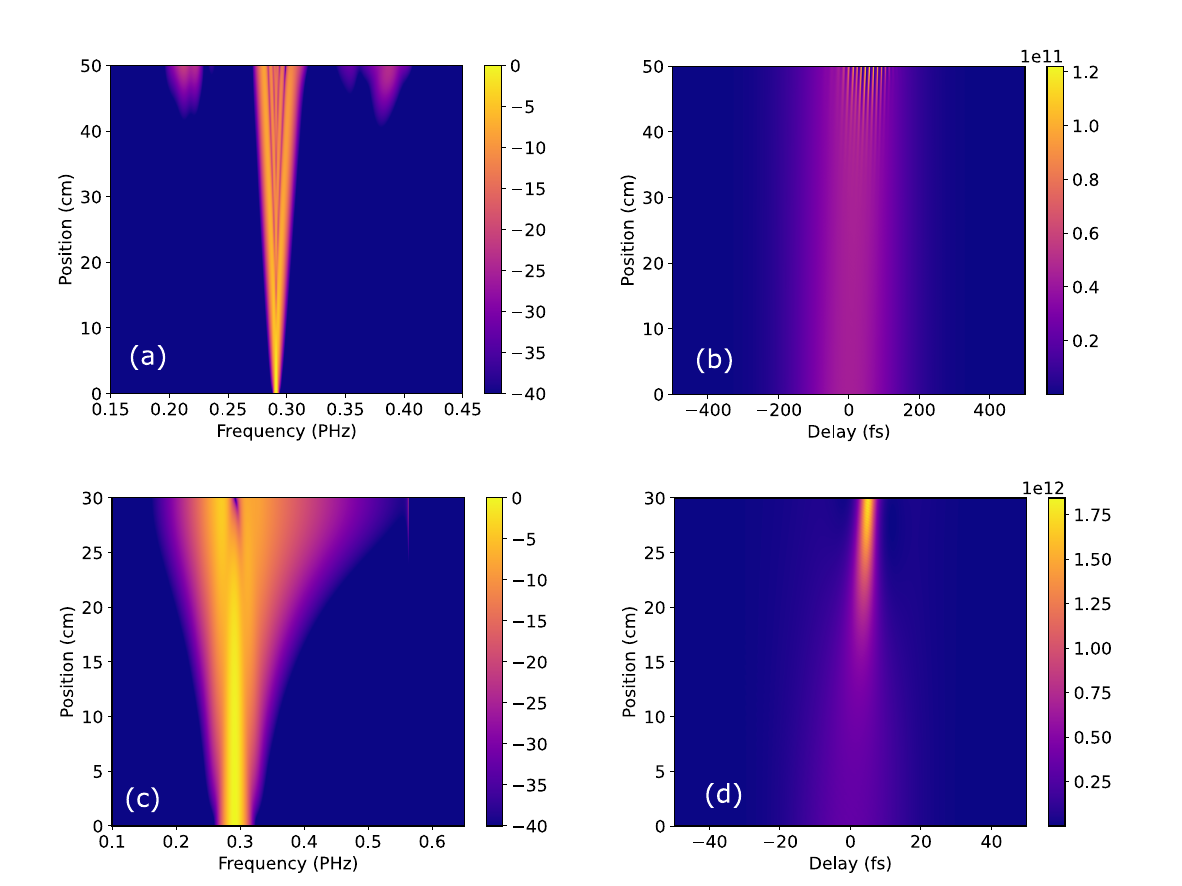}
  \caption{\textbf{Numerical simulation for nonlinear pulse propagation in the fiber stages.} (a) The spectral evolution of the first fiber stage based on self-phase modulation. (b) The corresponding time domain propagation. External methods are required for pulse compression to its Fourier transform. (c) Spectral evolution of the second fiber stage. (d) The time evolution indicates the temporal compression of the pulses to 3\,fs at the fiber output, due to soliton self-effect compression. These values correspond to 317\,TW/cm$^2$ peak intensity. The numerical simulations used a model described in \cite{tani2014multimode}.}
  \label{fig:FS}
    \end{center}
\end{figure}

\begin{figure}[h]
\begin{center}
  \includegraphics[width=1\linewidth]{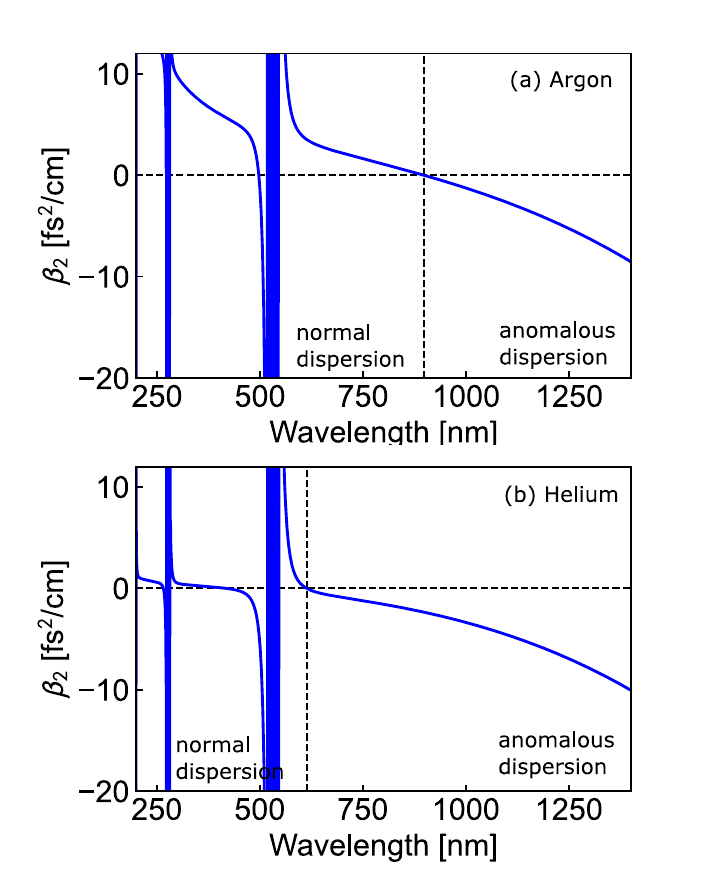}
  \caption{\textbf{Fiber dispersion tuning.} (a) Fiber dispersion for SR-PCF when filled with 15\,bar of Argon. (b) Fiber dispersion for SR-PCF when filled with 15\,bar of Helium. Both curves are calculated using the model described in this paper \cite{zeisberger2017analytic}. Two discontinuities stretching longitudinally indicate first and second-order resonances inside the fiber due to core-wall capillary thickness. The intersection of the blue curve and the dotted line indicates the zero dispersion wavelength (ZDW). The region to the right of ZDW has anomalous dispersion, whereas the region to the left has normal dispersion.}
  \label{fig:FD}
    \end{center}
\end{figure}

\clearpage
\subsection*{Field-resolved detection}
\begin{figure}[h]
\begin{center}
  \includegraphics[width=1\linewidth]{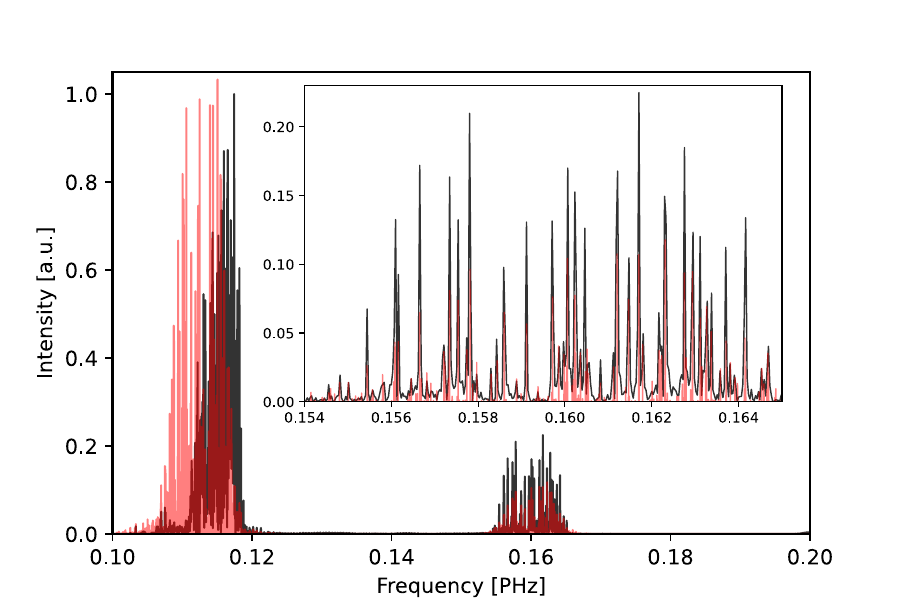}
  \caption{\textbf{Comparison with HITRAN.} The black spectra obtained from the field-resolved water measurement at 50\,\% relative humidity, shown in Figure 4(a), match the red HITRAN database spectra. The inset figure displays the combination band at 0.160 PHz, in agreement with the HITRAN spectrum. At 0.10\,PHz and 0.12\,PHz spectral range, the measured spectrum appears to be blue-sifted compared to the HITRAN reference spectrum, which is caused by the narrower bandwidth of our excitation spectrum.}
  \label{fig:HITRAN}
  \end{center}
\end{figure}

\begin{figure}[h]
\begin{center}
  \includegraphics[width=1\linewidth]{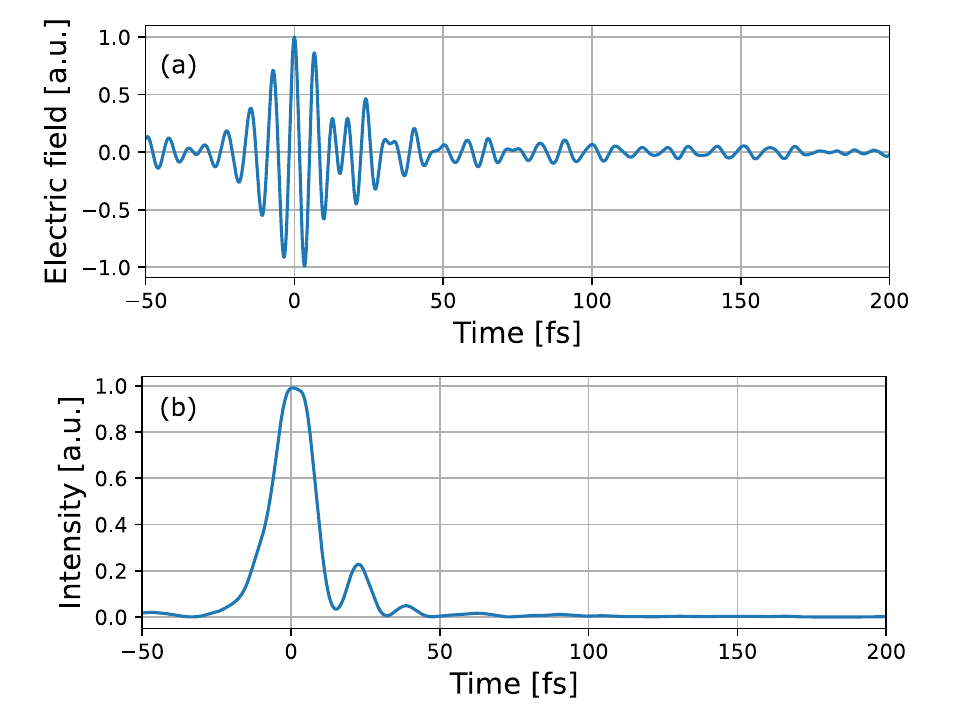}
  \caption{\textbf{Molecular information in the electric field vs intensity.}  (a) Electric field of the molecular response when excited impulsively. (b) The detected intensity counterpart in pump-probe intensity techniques. In field-resolved detection the measurement signal scales linearly with the electric field. Therefore molecular response is recorded by higher sensitivity than the pump-probe techniques.}
  \label{fig:IF}
    \end{center}
\end{figure}
\clearpage
\subsection*{Effectiveness of temporal gating in isolating discrete responses} \label{tempgate}
\begin{figure}[h]
\begin{center}
  \includegraphics[width=0.9\linewidth]{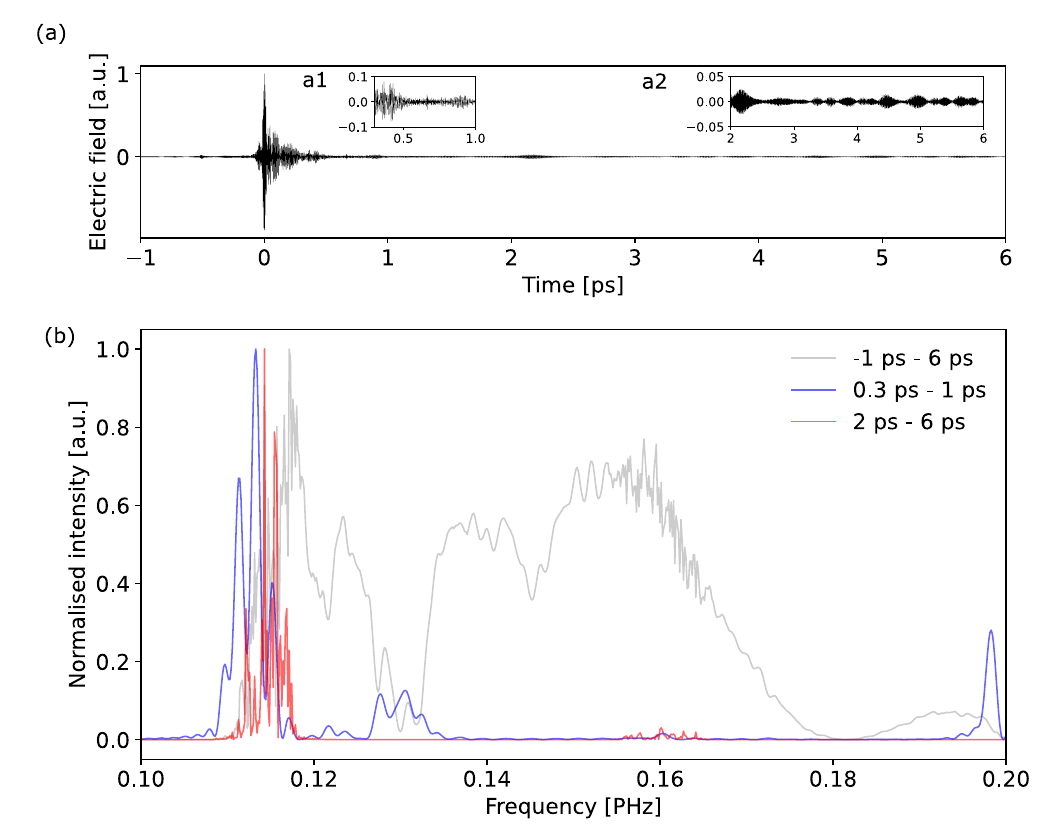}
  \caption{\textbf{Temporal gating in ethanol sample.} (a) The sampled electric field with 82.5\,$\mu$mol of liquid ethanol in the beam path. Inset a1 and a2 show two distinct temporal gating times. (b) The black color represents the Fourier transform of the entire waveform, while the blue and red colors denote the Fourier transform of the inset a1 and a2, respectively. The legend illustrates the time frame for which the Fourier transform is applied.}
  \label{fig:TG}
  \end{center}
\end{figure}
The dephasing duration of molecular response is subject to the molecular phase and varies from hundreds of femtoseconds in the liquid state to a few nanoseconds in the gas phase. Therefore, effective time filtering enables relative isolation of phase responses. Fig.\,\ref{fig:TG} (a) shows the full trace of the measured electric field for ethanol, while a1 and a2 in the inset show two magnified regions of the trace at the trailing edge of the femtosecond excitation pulse corresponding to the short-lived liquid response, and long-lived gas response, respectively. The Fourier transform of different temporally gated areas of panel (a) is shown in panel (b). The black spectrum shows the Fourier Transform of the entire sampled field, while the blue spectrum shows the absorption peaks of ethanol corresponding to panel a1, and the red spectrum shows the absorption of atmospheric water vapor corresponding to Figure a2. The absorption band for the O-H stretch in ethanol and water overlaps.

\clearpage

\begin{figure}[h]
\begin{center}
  \includegraphics[width=0.8\linewidth]{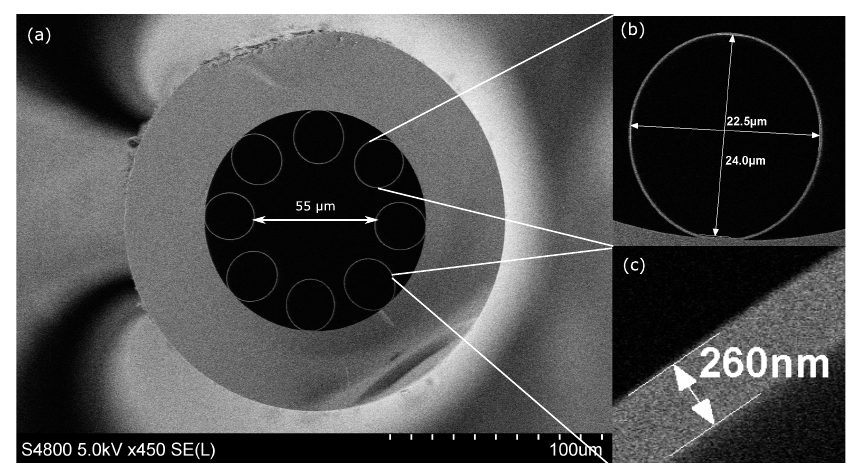}
  \caption{\textbf{SEM fiber images.} Both nonlinear fiber stages use an SR-PCF. (a) The SEM image shows a 55\,$\mu$m diameter fiber with 8 capillaries. (b) Shows the capillary diameter of 23\,$\mu$m.  (c) The core-wall thickness of the capillary is 260\,nm. }
  \label{fig:SEM}
  \end{center}
\end{figure}

\begin{figure}[h]
  \includegraphics[width=0.9\linewidth]{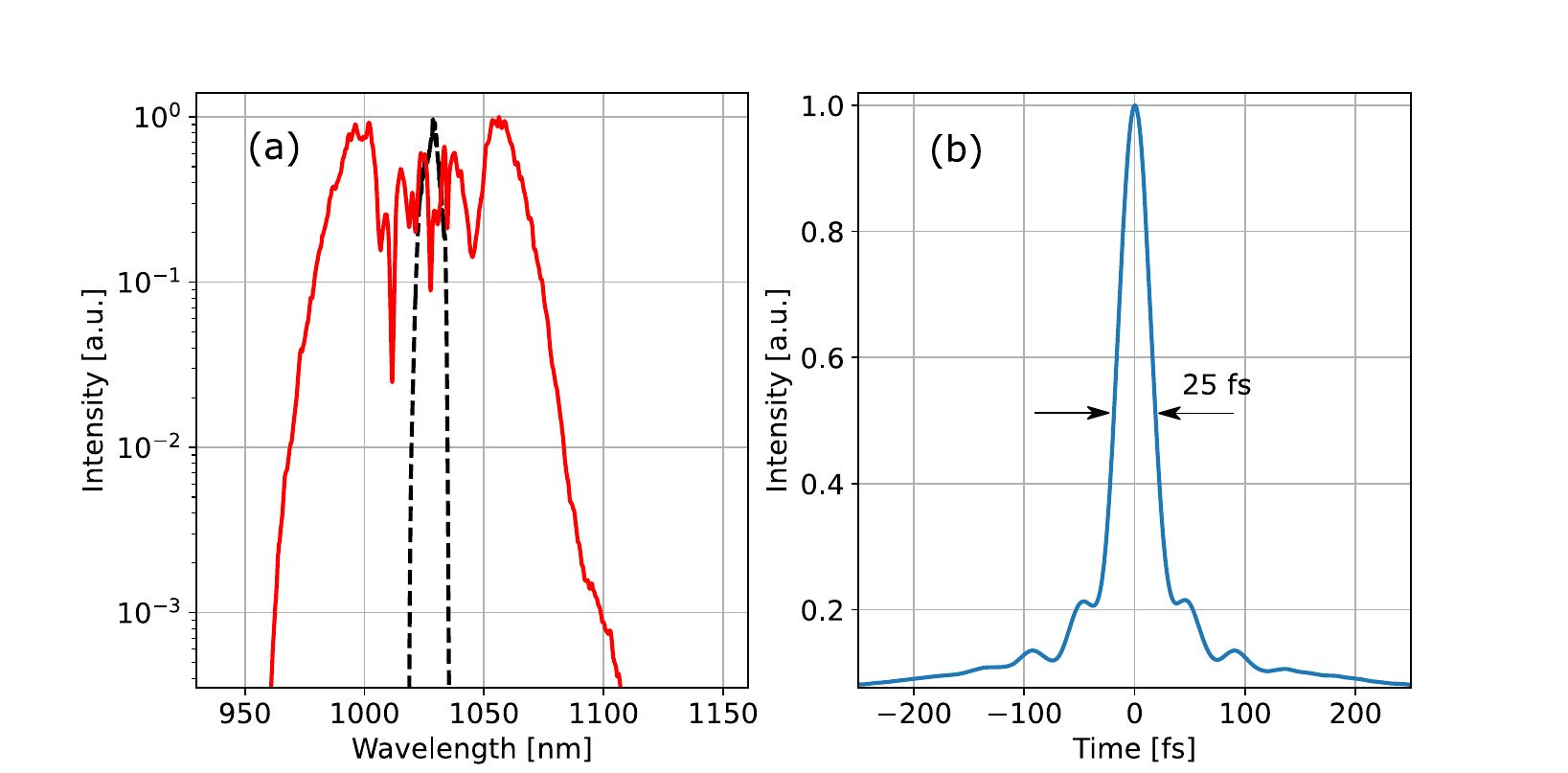}
  \begin{center}
  \caption{\textbf{First fiber stage characterization.} (a) The first nonlinear fiber stage's output spectrum (red), and the dashed line represent the input laser spectrum. (b) The autocorrelation trace of the same stage was measured at the input to the second stage, and the pulse duration at the FWHM was found to be 25\,fs.}
  \label{fig:S1}
  \end{center}
\end{figure}

\begin{figure}[h]
\begin{center}
  \includegraphics[width=0.7\linewidth]{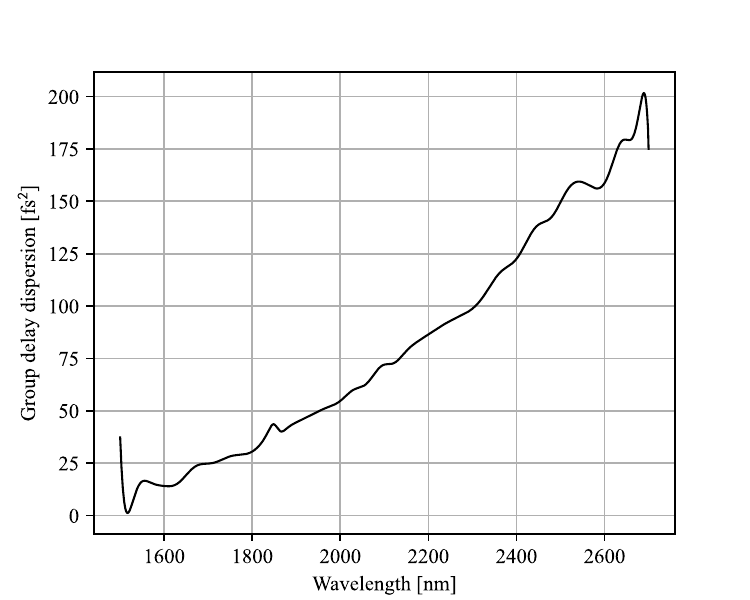}
  \caption{\textbf{UFI IR7202 dispersion curve.} Group delay dispersion of a pair of double-angle chirp mirrors used for NIR pulse compression. In total a pair of 3 double-angled chirp mirrors were used to achieve the compression.}
  \label{fig:CM}
    \end{center}
\end{figure}

\begin{figure}[h]
\begin{center}
  \includegraphics[width=0.7\linewidth]{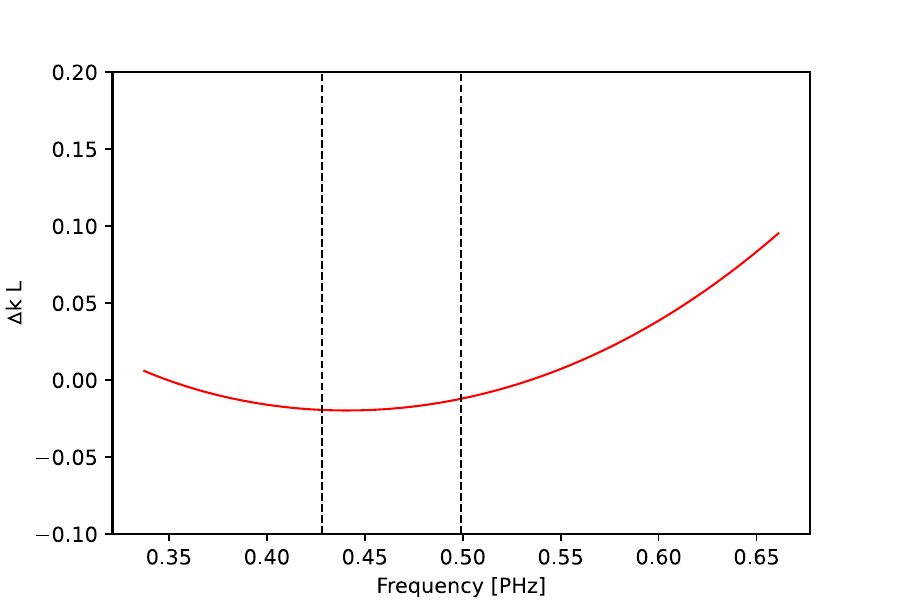}
  \caption{\textbf{EOS phase matching bandwidth.} Calculated the wavevector mismatch ($\Delta$k) for L = 10\,$\mu$m BBO, which is related to the EOS phase-matching bandwidth. The detection bandwidth of the EOS is indicated between two dashed lines, which shows a nearly flat response. The numerical calculations were performed using SISYFOS \cite{arisholm1997general}.}
  \label{fig:MM}
  \end{center}
\end{figure}

\clearpage 

\bibliography{sn-bibliography.bib}% common bib file

\end{document}